# Spontaneous mechanical oscillation of a DC driven single crystal

Kim L. Phan[*], Peter G. Steeneken[*], Martijn J. Goossens, Gerhard E.J. Koops, Greja J.A.M. Verheijden, and Joost T.M. van Beek

*NXP-TSMC Research Center, NXP Semiconductors, HTC 4, 5656 AE Eindhoven, the Netherlands*

[*]*These authors contributed equally to this work*

**There is a large interest to decrease the size of mechanical oscillators[1-15] since this can lead to miniaturization of timing and frequency referencing devices[7-15], but also because of the potential of small mechanical oscillators as extremely sensitive sensors[1-6]. Here we show that a single crystal silicon resonator structure spontaneously starts to oscillate when driven by a constant direct current (DC). The mechanical oscillation is sustained by an electrothermomechanical feedback effect in a nanobeam, which operates as a mechanical displacement amplifier. The displacement of the resonator mass is amplified, because it modulates the resistive heating power in the nanobeam via the piezoresistive effect, which results in a temperature variation that causes a thermal expansion feedback-force from the nanobeam on the resonator mass. This self-amplification effect can occur in almost any conducting material, but is particularly effective when the current density and mechanical stress are concentrated in beams of nano-scale dimensions.**

To demonstrate the effect, experiments are presented which show that a single crystal resonator of *n*-type silicon spontaneously starts to oscillate at a frequency of 1.26 MHz when the applied DC current density in the nanobeam exceeds 2.83 GA/m$^2$. The homogeneous monolithic oscillator device, which is shown in figures 1a and 1b, consists of a mass measuring 12.5×60.0×1.5 μm$^3$, which is suspended by a 3 μm wide spring beam and a 280 nm narrow nanobeam. The structure is made from a 1.5 μm thick phosphor-doped silicon layer on a 150 mm diameter silicon-on-insulator (SOI) wafer, with a phosphor doping concentration $N_d$=4.5×10$^{18}$ cm$^{-3}$ giving a specific resistivity of 10$^{-4}$ Ωm. The thin crystalline silicon layer is structured in a single mask step by a deep reactive ion etch (DRIE) and the buried SiO$_2$ layer below the mass and beams is removed in a hydrogen fluoride vapour etch, with an underetch distance of 7 μm. The in-plane mechanical bending resonance mode that determines the oscillation frequency is indicated by an arrow in figure 1a. A DC current $I_{dc}$ can be driven through the beams between terminals T1 and T2. As a result of the geometry of the structure, the current density and mechanical stress are concentrated in the narrow nanobeam, but the resonance frequency of the resonator is mainly determined by the spring beam and resonator mass. Measurements are performed at room temperature in a vacuum chamber at low pressure to reduce the effects of gas damping.

Before discussing the stand-alone operation of the oscillator, its in-plane mechanical resonance at 1.26 MHz is first characterized by actuating it with an AC electrostatic force on terminal T3 and detecting the displacement via the piezoresistive effect using the method described in reference 16. The resonant displacement, which is proportional to the transconductance $g_m$, is shown in figure 2a.

Now, the external AC voltage is disconnected from terminal T3. Thus, as shown in figure 1a, the device is only connected at terminal T1 to a DC current source $I_{dc}$ and to an oscilloscope. All other terminals are grounded. At low values of $I_{dc}$ only noise is observed on the oscilloscope. However, if the DC current is increased above a threshold $I_{osc}$=1.19 mA a remarkable effect occurs: the device spontaneously starts to oscillate and generates an AC output signal $v_{ac}$ with a frequency of 1.26 MHz as is shown in figures 2b, 2c and 2d.

This spontaneous oscillation of the mechanical resonator is a result of the self-amplification of its motion by the nanobeam. This amplification occurs via the following electrothermomechanical feedback effect, which is also illustrated by the diagram in figure 3a. Suppose the mass is moving in a direction that compresses the nanobeam. Due to the negative piezoresistive gauge factor of the *n*-type silicon, the compressive strain in the nanobeam causes an increasing resistance via the piezoresistive effect. This increases the resistive heating power in the nanobeam, which results in an increasing temperature, after a thermal delay. The temperature increase causes a thermal expansion force, which acts as a feedback force on the mass. The nanobeam therefore acts as a mechanical feedback amplifier. At small DC currents its feedback mechanism is not strong enough to compensate for the intrinsic damping of the mechanical resonator. However, if $I_{dc}$ exceeds the oscillation threshold current $I_{osc}$ the device starts to oscillate and generates a sinusoidal output voltage $v_{ac}$. The modulation in resistance $r_{ac}$ in combination with $I_{dc}$ generates an AC voltage $v_{ac}$ across the device which can be measured at the output terminal (T1) of the device as shown in figures 2b, 2c and 2d.

Figure 2b shows the start-up of the oscillator at $I_{dc}$=$I_{osc}$+0.01 mA=1.20 mA. It was observed that the start-up time can be significantly reduced by increasing $I_{dc}$. After start-up, as shown in figure 2c, the oscillator generates a stable sinusoidal voltage $v_{ac}$ at a DC power $P_{dc}$=1.19 mW. The spectrum of the signal shown in figure 2d shows that the power spectral density (*PSD*) of the noise is -70 dBc/Hz at 10 Hz from the carrier frequency. Because the total noise power is the sum of amplitude and phase noise, the phase noise can be even lower. The noise floor is expected to be mainly determined by the resistive noise of the internal resistance $R_{dc}$=824 Ω. The calculated resistive noise power spectral density is $10 \cdot \log[(4k_B T R_{dc})/v_{ac,rms}^2]$=-134 dBc/Hz at room temperature, where $k_B$ is Boltzmann's constant, the temperature of the nanobeam $T \approx 300$ K and the r.m.s. voltage $v_{ac,rms}$=19 mV is determined from figure 2c. This estimate of the resistive noise compares well to the noise floor in preliminary phase noise measurements of the device. To investigate the robustness of the oscillation, a sample of 12 mechanical resonators on the wafer is tested. All 12 devices oscillate at a pressure of 0.01 mbar and their threshold current has a small spread of $I_{osc}$=(1.21±0.03) mA. The oscillation is also observed in devices with different geometries on different wafers, but gives the strongest signal for the geometry shown in figure 1b.

To quantitatively analyze the operation mechanism of the oscillator, the device is represented by a simplified small-signal electrical model, which is shown in figure 3b. The linearized differential equations in the different physical domains are represented by this electrical circuit, which is discussed in more detail in Supplementary Discussion 1. The circuit elements in the mechanical, electrical and thermal domains are separated by dashed lines. The mechanical harmonic oscillator is represented by an equivalent *RLC* network, in which the component values are given in terms of the mechanical mass $m$, spring constant $k$ and damping coefficient $b$, by $L_m=m$, $C_m=1/k$ and $R_m=b$. The undamped resonance frequency of the lowest in-plane bending mode of the resonator is $\omega_0=\sqrt{(k/m)}$ and its intrinsic $Q$-factor is $Q_{int}=m\omega_0/b$. The displacement of the centre-of-mass $x$ causes a strain in the spring, which generates a resistance change $r_{ac}=K_{pr}R_{dc}x$ via the piezoresistive effect, where $K_{pr}$ is defined as the effective piezoresistive gauge factor. This will generate an AC voltage $v_{ac}$ across the nanobeam, which is represented by a voltage-controlled voltage source with output voltage $v_{pr}=I_{dc}r_{ac}$ in figure 3b. The AC resistive heating power in the beam is given by $p_{ac}=I_{dc}v_{ac}$ and is represented by the current $i_t$ from a voltage-controlled current source. The generated thermal power is partially stored in the heat capacitance and partially leaks away through the thermal conductance of the beam, which are represented by the capacitor $C_t$ and resistor $R_t$. This power causes a temperature change $T_{ac}$, which results in a thermal expansion force $F_{te}=k\alpha T_{ac}$ that is represented by a voltage-controlled voltage source, where $\alpha$ is defined as the effective thermal expansion coefficient.

The thermal expansion force in the network of figure 3b is found to be $F_{te}=\alpha k I_{dc}^2 R_{dc} K_{pr} x/(1/R_t+i\omega C_t)$, by multiplication of the transfer functions in the feedback loop. By substituting this feedback force in the equation for the damped, driven harmonic mechanical oscillator, as shown in Supplementary Discussion 2, the mechanical damping force $F_{damp}=-b \cdot dx/dt$ is exactly cancelled by the feedback force $F_{te}$ at a threshold value of the DC current $I_{dc}=I_{osc}$, which is given by the following equation:

$$\frac{1}{Q_{int}} = I_{osc}^2 \, \text{Im} \, \beta$$

$$\text{with } \beta = \frac{\alpha R_{dc} K_{pr}}{(1/R_t + i\omega_0 C_t)} \tag{1}$$

For values of $I_{dc}>I_{osc}$ the power gain from the feedback force becomes larger than the intrinsic mechanical loss of the resonator. Therefore the amplitude of the oscillation will increase in time until it is limited by non-linear effects which stabilize the sustained output signal. Since equation (1) can only be met if the imaginary part of $\beta$ is positive, the thermal delay caused by the heat capacitor $C_t$ and the negative value of $K_{pr}$ are important to meet the oscillation condition.

The threshold DC current $I_{osc}$ needed to bring the device into oscillation is detected at different chamber pressures with an oscilloscope. Care was taken to keep the impedance of the detection circuit high compared to $R_{dc}$. In figure 4 the values of $1/Q_{int}$, which are obtained from the fits of the transconductance $g_m$ curves in figure 2a, are plotted as a function of the square of this threshold current $I_{osc}^2$. The data closely follow a straight line through the origin as predicted by equation (1), with a fitted slope Im $\beta$=50 A$^{-2}$. At large currents the data deviate from the linear fit, possibly because of non-linear effects or by the temperature or pressure dependence of the device parameters.

From finite element method (FEM) simulations discussed in the Supplementary Discussions 3 and 4 the thermal parameters of the device are estimated to be $\alpha$=43.1×10$^{-12}$ m/K, $R_t$=6.74×10$^3$ K/W and $C_t$=9.29×10$^{-12}$ J/K. By substituting these simulated values and the measured values of $\omega_0$, $R_{dc}$ and $K_{pr}$ in equation (1), a value of Im $\beta$=63 A$^{-2}$ is found. A direct FEM calculation from the geometry and the literature values of the silicon material parameters yields Im $\beta$=58 A$^{-2}$ (see Supplementary Discussion 3). This quantitative agreement between the measured value Im $\beta$=50 A$^{-2}$ from figure 4 and the values determined from equation (1) and by FEM simulations, support the validity of the proposed oscillator model.

Although the presented feedback mechanism is strong in *n*-type silicon, due to its large negative piezoresistive coefficients[17], the mechanism plays a role in almost any conducting material and might therefore also be used to create oscillators of different composition. Devices made out of materials with positive piezoresistive coefficients, in which the sign of $K_{pr}$ is reversed, can also be brought into oscillation, by replacing the DC current source $I_{dc}$ with a constant DC voltage source. When connected to a voltage source the sign of the AC resistive heating power $p_{ac}$ is reversed, such that Im $\beta$ is still positive and condition (1) can be met.

The self-amplification mechanism, which spontaneously brings a homogeneous mechanical resonator into sustained oscillation from a constant DC current flow, is a result of the intrinsic material properties and geometry of the resonator. Related self-amplification mechanisms by which a mechanical resonator can be brought into sustained oscillation from a constant flow are aeroelastic flutter in a steady gas or fluid flow, and mechanical oscillations as a result of a constant optical radiation pressure[18,19].

Besides its scientifically relevant oscillation mechanism, the oscillator also shows interesting application prospects. Because it does not require additional transistor-amplifiers or transducers, the oscillator structure can be manufactured in standard semiconductor technologies using a single mask step process. Because of its low noise and low power consumption, it is suitable for clocks, frequency synthesizers and actuator systems. Moreover, the presented mechanism can enable further oscillator miniaturization by using carbon nanotubes or silicon nanowires with giant piezoresistance effect[20-22] as feedback element. Such nanomechanical oscillators can be made extremely sensitive to variations in mass and force, thus enabling them to be applied in physical, chemical and biological sensors and sensor arrays.

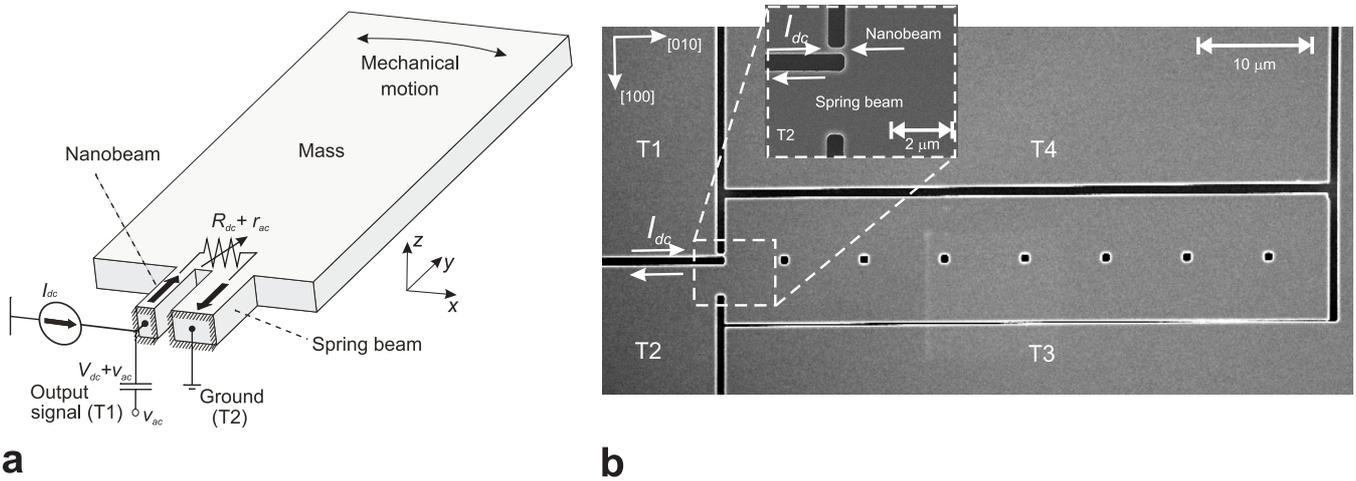

**Figure 1** Drawing and micrographs of the oscillator and its terminals T1-T4. **a**, Drawing of the mechanical oscillator (not to scale) and its signal (T1) and ground (T2) terminals. To operate the oscillator a DC current $I_{dc}$ is applied between these terminals and the output voltage $v_{ac}$ is measured. **b**, Top-view image of the device made with a Scanning-Electron-Microscope (SEM). The inset shows a magnification of the wide spring and narrow nanobeam by which the proof mass is suspended. Holes in the mass facilitate the etching of buried oxide below the structure.

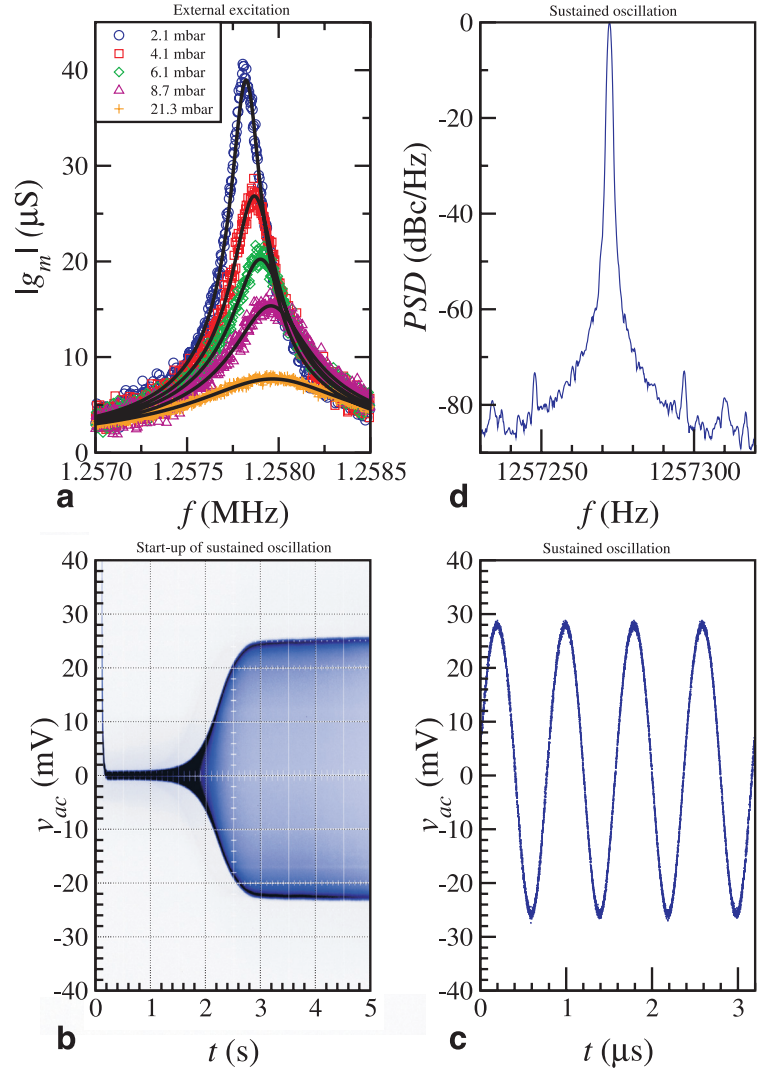

**Figure 2** Oscillator output. **a**, Characterization of the in-plane mechanical bending resonance by external excitation at different chamber pressures. The resonance is excited by an AC electrostatic force generated by a voltage $V_{act,dc}+v_{act,ac}$ on terminal T3, with $V_{act,dc}= -1$ V. The displacement is detected via the piezoresistive effect, which can be measured using a small probe current $I_{dc}$=0.1 mA. From this measurement the transconductance $g_m$ is determined[16], which is proportional to the displacement. Solid black lines are fits of the data that are used to determine $K_{pr}$=-6.4×10$^5$ m$^{-1}$ and the pressure dependence of $Q_{int}$ as discussed in Supplementary Discussion 4. **b,c,d**, Stand-alone operation of the oscillator at $I_{dc}=I_{osc}$+0.01 mA at a chamber pressure $P$=0.01 mbar. In panel **b** the spontaneous startup of the mechanical oscillations is measured by an analog oscilloscope when a current $I_{dc}$=1.20 mA is switched on at $t$=0 s. The sustained sinusoidal oscillator output signal is measured **c** by a digital oscilloscope and **d** by a spectrum analyzer that determines the power spectral density (*PSD*). From the oscilloscope data in figure 2c, the amplitude of the centre-of-mass is estimated to be $x_0$≈43 nm, using $K_{pr}x_0=v_{ac0}/V_{dc}$, with $V_{dc}=I_{dc}R_{dc}$=0.99 V and an AC amplitude $v_{ac0}$=27 mV.

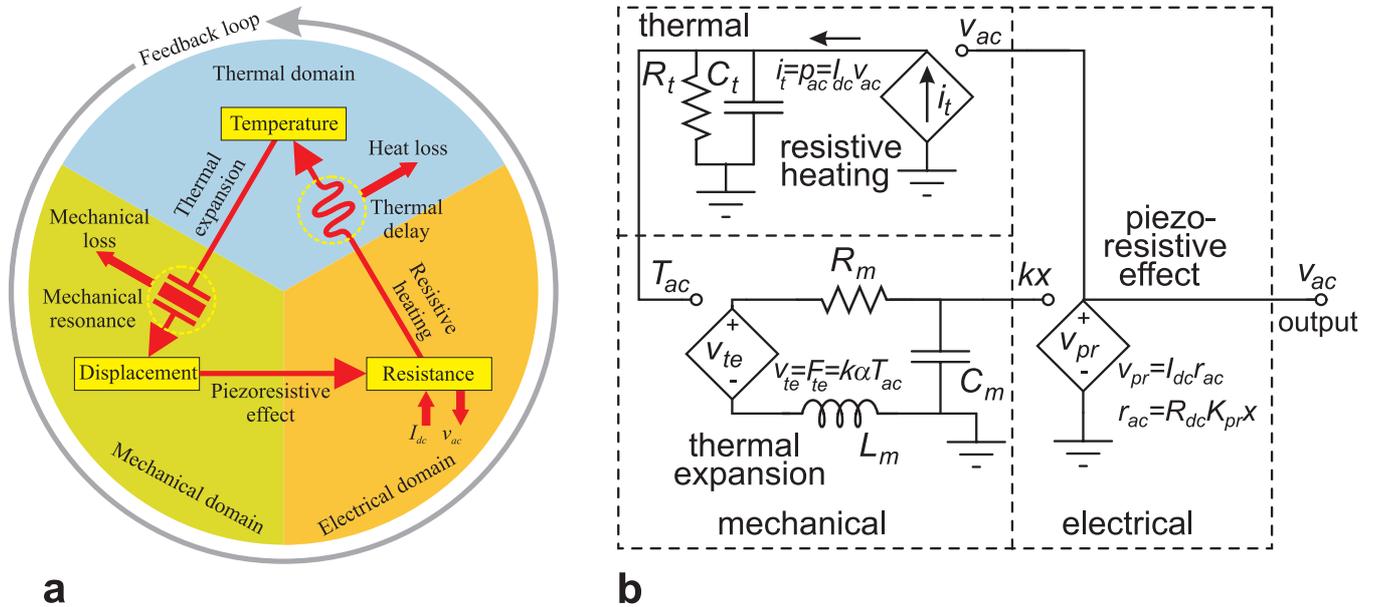

**Figure 3** Schematic operation mechanism. **a,** Diagram, qualitatively showing the feedback loop with the transduction mechanisms which connect the three physical quantities: displacement $x$, temperature $T_{ac}$ and voltage $v_{ac}$. Also shown are the high-$Q$ mechanical resonator and the thermal delay that causes a phase shift. **b,** Small-signal AC equivalent circuit of the oscillator, where the thermal, mechanical and electrical differential equations are represented by a linearized electrical circuit. Dashed lines separate the three physical domains. Transduction mechanisms are represented by controlled current and voltage sources.

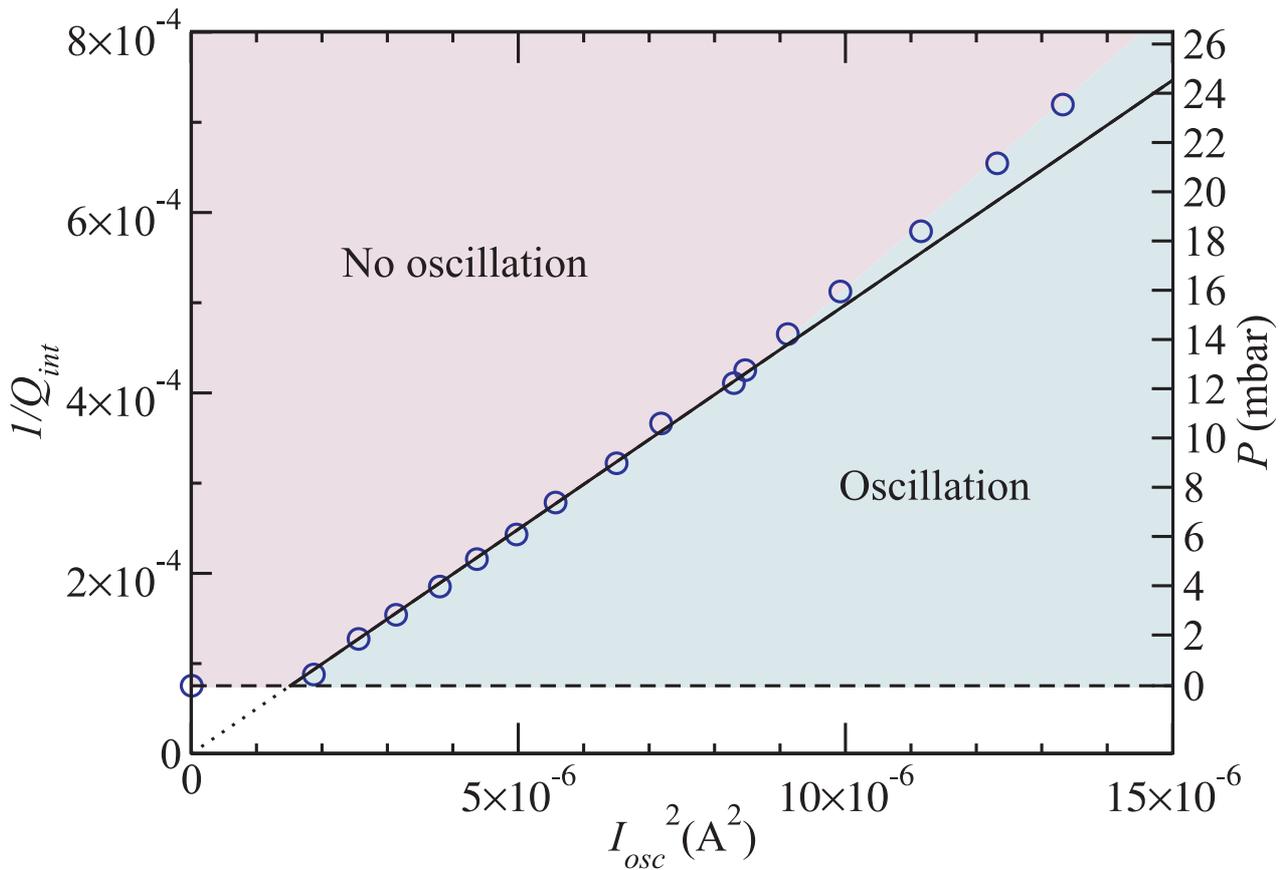

**Figure 4** Measurement of the square of the oscillation threshold current $I_{osc}^2$ as a function of the inverse Q-factor $1/Q_{int}$. The solid line is a linear fit through the origin with a slope Im $\beta$=50 A$^{-2}$. The dashed line indicates the Q-factor in the absence of gas damping $Q_{int0}$=13.3×10$^3$, which prevails at chamber pressures $P$ below 0.01 mbar. The scale on the right y-axis shows the chamber pressure based on a linear fit of a $1/Q_{int}$ versus $P$ measurement. If $I_{dc}$ exceeds $I_{osc}$ the device will exhibit sustained oscillation as indicated by the colored area.

**References**

1. Sazonova, V. *et al.* A tunable carbon nanotube electromechanical oscillator. *Nature* **431**, 284-287 (2004).

2. Meyer, J. C., Paillet, M. and Roth, S., Single-molecule torsional pendulum. *Science* **309**, 1539-1541 (2005).

3. Feng, X. L., White, C. J., Hajimiri, A. and Roukes M. L., A self-sustaining ultrahigh-frequency nanoelectromechanical oscillator. *Nature Nanotech.* **3**, 342-346 (2008).

4. Jensen, K., Kim, K. and Zettl, A., An atomic-resolution nanomechanical mass sensor. *Nature Nanotech.* **3**, 533-537 (2008).

5. Bedair, S.S. and Fedder, G.K., CMOS MEMS oscillator for gas chemical detection. *Proc. IEEE Sensors,* 955-958 (2004).

6. Verd, J. *et al.* Monolithic CMOS MEMS oscillator circuit for sensing in the attogram range. *IEEE Electr. Dev. L.* **29**, 146-148 (2008).

7. Nathanson, H.C., Newell, W.E., Wickstrom, R.A. and Davis, J.R., Jr., The resonant gate transistor. *IEEE T. Electron Dev.* **14**, 117 – 133 (1967).

8. Wilfinger, R. J., Bardell, P. H. and Chhabra, D. S., The resonistor: a frequency selective device utilizing the mechanical resonance of a silicon substrate. *IBM J. Res. Dev.* **12**, 113-117 (1968).

9. Nguyen, C.T.-C., Howe, R.T., CMOS micromechanical resonator oscillator. *Proc. IEEE Electron. Dev. Mtg. (IEDM)*, 199-202 (1993).

10. Reichenbach, R.B., Zalalutdinov, M., Parpia, J.M. and Craighead, H.G., RF MEMS oscillator with integrated resistive transduction. *IEEE Electr. Dev. L.* **27**, 805-807 (2006).

11. van Beek, J.T.M. *et al.* Scalable 1.1 GHz fundamental mode piezo-resisitive silicon MEMS resonators. *Proc. IEEE Electron. Dev. Mtg. (IEDM)*, 411-414 (2007).

12. Lutz, M. *et al.* MEMS Oscillators for high volume commercial applications. *Proc. Transducers '07,* 49-52 (2007).

13. Grogg, D., Mazza, M., Tsamados, D. and Ionescu, A.M., Multi-gate vibrating-body field effect transistor (VB-FETs). *Proc. IEEE Electron. Dev. Mtg. (IEDM)*, 1-4 (2008).

14. Rantakari, P. *et al.* Low noise, low power micromechanical oscillator. *Proc. Transducers '05,* 2135 - 2138 (2005).

15. Lin, Y-W. *et al.* Series-resonant VHF micromechanical resonator reference oscillators. *IEEE J. Solid-St. Circ.* **39**, 2477- 2491 (2004).

16. van Beek, J.T.M., Steeneken, P.G. and Giesbers, B., A 10 MHz Piezoresistive MEMS resonator with high-Q. *Proc. Int. Freq. Contr. Symp.*, 475-480 (2006).

17. Smith, C. S., Piezoresistance effect in germanium and silicon. *Phys. Rev.* **94**, 42-49 (1954).

18. Rokhsari, H., Kippenberg, T.J., Carmon, T. and Vahala, K. J., Radiation-pressure-driven micro-mechanical oscillator. *Opt. Express* **13**, 5293-5301 (2005).

19. Kippenberg, T.J., Rokhsari, H., Carmon, T., Scherer, A. and Vahala, K. J., Analysis of radiation-pressure induced mechanical oscillation of an optical microcavity. *Phys. Rev. Lett.* **95**, 033901 (2005).

20. He, R. and Yang, P., Giant piezoresistance effect in silicon nanowires. *Nature Nanotech.* **1**, 42-46 (2006).

21. Tombler, T.W. *et al.* Reversible electromechanical characteristics of carbon nanotubes under local-probe manipulation. *Nature* **405**, 769-772 (2000).

22. Cao, J., Wang, Q., and Dai, H., Electromechanical properties of metallic, quasimetallic, and semiconducting carbon nanotubes under stretching. *Phys. Rev. Lett.*, **90**, 157601 (2003).

**Acknowledgements** We thank J.J.M. Ruigrok, C.S. Vaucher, K. Reimann, C. v.d. Avoort, R. Woltjer and E.P.A.M. Bakkers for discussions and suggestions and thank J. v. Wingerden for his assistance with the SEM measurements.

**Author Contributions**: K.L.P., J.T.M.v.B., P.G.S. and M.J.G. invented and designed the oscillator, P.G.S. and K.L.P. wrote the paper and performed the measurements, P.G.S. performed the FEM simulations and the quantitative analysis of the feedback mechanism, J.T.M.v.B., G.E.J.K. and G.J.A.M.V. developed the process technology and manufactured the device.

Supplementary Information (SI) to accompany

# Spontaneous mechanical oscillation of a DC driven single crystal

Kim L. Phan[*], Peter G. Steeneken[*], Martijn J. Goossens, Gerhard E.J. Koops, Greja J.A.M. Verheijden, and Joost T.M. van Beek

**The Supplementary Information contains the following sections:**

**I. Supplementary Discussions**

- Supplementary Discussion 1: The equations on which figure S1 is based are discussed.

- Supplementary Discussion 2: Equation (S1), which is identical to equation (1) from the main paper, is derived from the small signal circuit in figure S1:

$$\frac{1}{Q_{int}} = I_{osc}^2 \, \text{Im}\,\beta, \text{ with } \beta = \frac{\alpha R_{dc} K_{pr}}{(1/R_t + i\omega_0 C_t)} \quad (S1)$$

- Supplementary Discussion 3: The FEM model is discussed and is used to calculate Im $\beta$ directly from the geometry and material parameters.

- Supplementary Discussion 4: The equivalent circuit parameters shown in figure S1 are extracted from FEM simulations and measurements. The extracted values are compared to measurements and are substituted in equation (S1) to determine Im $\beta$ in an alternative way.

**II. Supplementary Figures**

- Supplementary Figure S1: Equivalent model small-signal circuit of the oscillator.

- Supplementary Figures S2-S5: FEM simulations of the oscillator.

**III. Supplementary Tables**

- Supplementary Table S1: FEM material parameters.

- Supplementary Table S2: Simulated and measured oscillator parameters.

**IV. Supplementary References [S1-S7]**

## I. Supplementary Discussions

**Supplementary Discussion 1: Equations on which figure S1 is based.**

*a. Mechanical domain*

In the mechanical domain, the resonance of the device is described by the driven damped harmonic oscillator equation. The mass *m*, damper *b* and spring *k* in this equation, are represented in figure S1 by an equivalent inductor $L_m=m$, resistor $R_m=b$ and capacitor $C_m=1/k$:

$$m\ddot{x} + b\dot{x} + kx = F_{te} = k\alpha T_{ac}$$
$$L_m\ddot{q}_m + R_m\dot{q}_m + \frac{q_m}{C_m} = k\alpha T_{ac} \qquad (S2)$$

This equation shows that the charge $q_m$ on the capacitor in the electrical circuit can represent the position *x* of the centre-of-mass of the oscillator. The thermal expansion force generated by the transducer element is represented by a voltage-controlled voltage source with output voltage $F_{te}=k\alpha T_{ac}$. Note that the representation of the thermal expansion effect by a voltage-controlled voltage source neglects thermoelasticity effects. This is a good approximation because the thermoelastic power in the thermal domain is much smaller than the power leaking away through the thermal resistor $R_t$ and because thermoelastic damping of the mechanical system is taken into account in the mechanical damping resistor $R_m$.

*b. Electrical domain*

As a result of the piezoresistive effect, the extension of the spring causes a change in resistance $r_{ac}$. This is described by the equation $r_{ac}/R_{dc}=K_{pr}x$, in which $R_{dc}$ is the DC resistance of the device and $K_{pr}$ is defined as the effective piezoresistive gauge factor. Because there is a DC current $I_{dc}$ running through the resistor, the piezoresistive effect can be represented by a voltage-controlled voltage source, with an output voltage $v_{pr}=I_{dc}r_{ac}$. This AC voltage $v_{ac}=v_{pr}$ can be detected at the output of the device. Note that for the circuit in figure S1 to be valid, the AC-impedance to ground at the output terminal should be high compared to $R_{dc}$. Otherwise, the resistive heating power $p_{ac}$ will be different, which can influence the feedback loop of the oscillator.

*c. Thermal domain*

The change in the resistance of the beam also results in a modulation of the resistive heating power $p_{ac}=I_{dc}^2 r_{ac}=I_{dc} v_{ac}$, which is represented by a voltage-controlled current source with output current $i_t=p_{ac}$ in figure S1. If the temperature of the nanobeam $T_{ac}$ is represented by an equivalent voltage, the thermal physics can be approximated by an electrical equation with a heat capacitance given by $C_t$ and a thermal resistance $R_t$:

$$C_t \frac{dT_{ac}}{dt} + \frac{T_{ac}}{R_t} = p_{ac} = i_t \tag{S3}$$

The thermal expansion force generated by $T_{ac}$ closes the feedback loop.

**Supplementary Discussion 2: Derivation of equation (S1) from figure S1.**

By multiplying the transfer functions in figure S1, the thermal feedback force can be expressed in terms of the displacement: $F_{te}=\alpha k I_{dc}^2 R_{dc} K_{pr} x/(1/R_t+i\omega C_t)$. To derive equation (S1) this feedback force is substituted in equation (S2). If the time dependence of the displacement is approximated by $x(t)=x_0 e^{i\omega t}$ and it is assumed that $\omega \approx \omega_0$ we find:

$$\begin{aligned}
&m\ddot{x} + b\dot{x} + kx = F_{te} \\
&\ddot{x} + \omega_0 \dot{x}/Q_{int} + \omega_0^2 x - F_{te}/m = 0 \\
&\left(-\omega^2 + i\omega\omega_0(1/Q_{int} - I_{dc}^2 \operatorname{Im}\beta) + \omega_0^2(1 - I_{dc}^2 \operatorname{Re}\beta)\right)x = 0 \\
&\beta = \frac{\alpha R_{dc} K_{pr}}{(1/R_t + i\omega_0 C_t)}
\end{aligned} \tag{S4}$$

This equation shows that the differential equation governing the linear system with feedback-force $F_{te}$ is identical to that of the damped undriven harmonic oscillator with a modified *Q*-factor $Q_{eff}$, with $1/Q_{eff}=1/Q_{int}-I_{dc}^2 \operatorname{Im}\beta$. For the device discussed in this paper, the value of Im β is positive. Therefore, $Q_{eff}$ will become infinite and the damping will become zero at a threshold value of the DC current $I_{dc}=I_{osc}$ given by:

$$1/Q_{int} = I_{osc}^2 \operatorname{Im}\beta \tag{S5}$$

This equation is identical to equation (1) in the main paper. Equation (S4) shows that for $I_{dc}>I_{osc}$ the effective damping becomes negative, which implies that the power gain from the feedback mechanism is larger than the power loss via the intrinsic mechanical damping. Therefore the amplitude of the oscillations will increase until it reaches a steady state sustained oscillation at a frequency $\omega_{osc}$ which is given by $\omega_{osc}^2 = \omega_0^2 (1 - I_{dc}^2 \text{Re}\,\beta)$. Because only the imaginary part of $\beta$ contributes to the increase in $Q_{eff}$, a phase shift in the feedback loop is needed for oscillation, which is provided by the thermal delay caused by the combination of the heat capacitor $C_t$ and resistor $R_t$ in figure S1.

**Supplementary Discussion 3: FEM model.**

*a. Linearization of FEM partial differential equations*

The state of the oscillator in the mechanical, electrical and thermal domains is described by the displacement **u**, voltage $V$ and temperature $T$. The partial differential equations (PDE) can be simplified by linearizing them around the DC bias point and by assuming that the variables all have a sinusoidal time dependence. The complex displacement **u**, voltage $V$ and temperature $T$ are then given by:

$$\begin{aligned}
\mathbf{u}(\mathbf{r},t) &= \mathbf{u}_{ac,r}(\mathbf{r})e^{i\omega t} \\
V(\mathbf{r},t) &= V_{dc}(\mathbf{r}) + v_{ac,r}(\mathbf{r})e^{i\omega t} \\
T(\mathbf{r},t) &= T_{dc}(\mathbf{r}) + T_{ac,r}(\mathbf{r})e^{i\omega t} + T_{RT}
\end{aligned} \quad (S6)$$

Where $T_{RT}$ is room temperature. As a result of the piezoresistive effect, the electrical conductivity $\sigma$ will also have an AC component:

$$\sigma(\mathbf{r},t) = \sigma_{dc} + \sigma_{ac}(\mathbf{r})e^{i\omega t} \quad (S7)$$

The amplitude of the displacement $\mathbf{u}_{ac,r}$ is assumed to be small. Therefore the voltage, temperature and resistance are small compared to the DC values:

$$\begin{aligned}
v_{ac,r}(\mathbf{r}) &\ll V_{dc}(\mathbf{r}) \\
T_{ac,r}(\mathbf{r}) &\ll T_{dc}(\mathbf{r}) \\
\sigma_{ac}(\mathbf{r}) &\ll \sigma_{dc}(\mathbf{r})
\end{aligned} \quad (S8)$$

Equation (S8) will be implicitly used to simplify the electrical and thermal PDEs shown below.

*b. Electrical equations*

The conductivity matrix σ relates the electric field **E** to the current density **J**:

$$\begin{aligned}(\sigma_{dc} + \sigma_{ac}e^{i\omega t})(\mathbf{E}_{dc} + \mathbf{E}_{ac}e^{i\omega t}) &= \mathbf{J}_{dc} + \mathbf{J}_{ac}e^{i\omega t} \\ \mathbf{J}_{dc} &= \sigma_{dc}\mathbf{E}_{dc} \\ \mathbf{J}_{ac} &= (\sigma_{dc}\mathbf{E}_{ac} + \mathbf{E}_{dc}\sigma_{ac})\end{aligned} \quad (S9)$$

Via the piezoresistive coefficients π [S1], the AC resistivity $\rho_{ac}$ depends on the mechanical stress **T**:

$$\rho_{ac} = \begin{bmatrix} \rho_{ac,11} \\ \rho_{ac,22} \\ \rho_{ac,33} \\ \rho_{ac,23} \\ \rho_{ac,31} \\ \rho_{ac,12} \end{bmatrix} = \rho_{dc}\begin{bmatrix} \pi_{11} & \pi_{12} & \pi_{12} & 0 & 0 & 0 \\ \pi_{12} & \pi_{11} & \pi_{12} & 0 & 0 & 0 \\ \pi_{12} & \pi_{12} & \pi_{11} & 0 & 0 & 0 \\ 0 & 0 & 0 & \pi_{44} & 0 & 0 \\ 0 & 0 & 0 & 0 & \pi_{44} & 0 \\ 0 & 0 & 0 & 0 & 0 & \pi_{44} \end{bmatrix}\begin{bmatrix} T_{11} \\ T_{22} \\ T_{33} \\ T_{23} \\ T_{31} \\ T_{12} \end{bmatrix} \quad (S10)$$

And the anisotropic electrical conductivity is the inverse of the resistivity matrix $\sigma = (\rho_{dc}+\rho_{ac}e^{i\omega t})^{-1}$:

$$\sigma_{dc} + \sigma_{ac}e^{i\omega t} = \frac{1}{\rho_{dc}^2}\left\{\begin{bmatrix} \rho_{dc} & 0 & 0 \\ 0 & \rho_{dc} & 0 \\ 0 & 0 & \rho_{dc} \end{bmatrix} - \begin{bmatrix} \rho_{ac,11} & \rho_{ac,12} & \rho_{ac,31} \\ \rho_{ac,12} & \rho_{ac,22} & \rho_{ac,23} \\ \rho_{ac,31} & \rho_{ac,23} & \rho_{ac,33} \end{bmatrix}e^{i\omega t}\right\} \quad (S11)$$

The electrostatic charge continuity equation is given by:

$$\nabla \cdot \mathbf{J} = \nabla \cdot \sigma\mathbf{E} = 0 \quad (S12)$$

Using equation (S9) the DC and AC electrical conduction equations are separated:

$$\nabla \cdot \mathbf{J}_{dc} = \nabla \cdot \sigma_{dc}\mathbf{E}_{dc} = 0 \quad (S13a)$$

$$\nabla \cdot \mathbf{J}_{ac} = \nabla \cdot (\sigma_{dc}\mathbf{E}_{ac} + \sigma_{ac}\mathbf{E}_{dc}) = 0 \quad (S13b)$$

The resistive heating power density $Q_t = Q_{dc} + Q_{ac}e^{i\omega t}$ is given by:

$$Q_{dc} = \mathbf{E}_{dc} \cdot \mathbf{J}_{dc} \quad (S14a)$$

$$Q_{ac} = \mathbf{J}_{dc} \cdot \mathbf{E}_{ac} + \mathbf{E}_{dc} \cdot \mathbf{J}_{ac} \quad (S14b)$$

*c. Thermal equations*

The equations governing the DC and AC thermal conduction are:

$$\nabla \cdot (k_h \nabla T_{dc}) + Q_{dc} = 0 \tag{S15a}$$

$$\nabla \cdot (k_h \nabla T_{ac,r} e^{i\omega t}) + Q_{ac} e^{i\omega t} = \rho_d c_p \frac{\partial T_{ac,r} e^{i\omega t}}{\partial t} = i\omega \rho_d c_p T_{ac,r} e^{i\omega t} \tag{S15b}$$

Here $k_h$ is the thermal conductivity, $c_p$ the specific heat capacity and $\rho_d$ the mass density. Convection is neglected since the device is operated in vacuum.

*d. Mechanical equations*

The mechanical partial differential equations consist of the equation of motion, the stress-strain relation and the strain-displacement relation including thermal expansion:

$$\begin{aligned} \nabla \cdot \mathbf{T} &= \rho_d \frac{\partial^2 \mathbf{u}_{ac,r}}{\partial t^2} \\ \mathbf{T} &= \mathbf{c}\mathbf{S} \\ \mathbf{S} &= \nabla_s \mathbf{u}_{ac,r} - \alpha_{vec} T_{ac,r} \end{aligned} \tag{S16}$$

**S** is the strain vector field, **T** is the stress vector field and **c** is the stiffness tensor, $\nabla_s$ is the symmetric-gradient operator representing the strain-displacement relation, **c** is the cubic anisotropic elasticity matrix of silicon and $\alpha_{vec}$ is the thermal expansion vector for a cubic material:

$$\mathbf{c} = \begin{bmatrix} c_{11} & c_{12} & c_{12} & 0 & 0 & 0 \\ c_{12} & c_{11} & c_{12} & 0 & 0 & 0 \\ c_{12} & c_{12} & c_{11} & 0 & 0 & 0 \\ 0 & 0 & 0 & c_{44} & 0 & 0 \\ 0 & 0 & 0 & 0 & c_{44} & 0 \\ 0 & 0 & 0 & 0 & 0 & c_{44} \end{bmatrix} \quad \alpha_{vec} = \begin{bmatrix} \alpha_t \\ \alpha_t \\ \alpha_t \\ 0 \\ 0 \\ 0 \end{bmatrix} \tag{S17}$$

*e. Boundary conditions*

The mechanical boundary conditions fix the structure ($\mathbf{u}_{ac,r}$=0 m) at the anchors. The thermal boundary conditions impose $T_{dc}$=0 K and $T_{ac,r}$=0 K at the anchors and thermal insulation on all other boundaries for both the AC and DC domain. The DC electrical boundary conditions impose $V_{dc}=I_{dc}R_{dc}$ at one anchor and $V_{dc}$=0 V at the other anchor. Because the device is connected to a current source the AC electrical boundary conditions impose that the AC current density normal to the boundaries is zero ($\mathbf{n}\cdot\mathbf{J}_{ac}$=0 A/m$^2$).

*f. FEM simulation*

To solve the model, the partial differential equations discussed above have been incorporated in a coupled model in Comsol Multiphysics 3.4 [S2]. The material parameters were taken from literature and are shown in table S1. The geometry was taken from the mask design and the dimensions of the nanobeam were measured more accurately using a scanning electron microscope. The measured sacrificial layer underetch distance of 7 μm is included in the geometry of the anchors. To increase the simulation speed and accuracy, the device was simulated in 2-dimensions by using the plane-strain approximation. First the DC electrical equation (S13a) is solved. This gives $\mathbf{E}_{dc}$ and $\mathbf{J}_{dc}$ which are needed in equations (S13b) and (S14b). Then the coupled AC mechanical-thermal-electrical eigenvalue equations (S13b),(S15b) and (S16) are solved, using the additional equations (S10), (S11), (S14b) and (S17). This yields the eigenvectors of the lowest bending mode: $\mathbf{u}_{ac,r}(\mathbf{r})$, $v_{ac,r}(\mathbf{r})$ and $T_{ac,r}(\mathbf{r})$. The solution also gives the complex angular eigenfrequency $\omega_{res}$ from which the effective $Q$-factor in the absence of mechanical damping $Q_{eff0}$ of the eigenmode can be calculated [S3] using the equation:

$$\frac{1}{Q_{eff0}} = \frac{2\operatorname{Im}\omega_{res}}{\operatorname{Re}\omega_{res}} = -I_{dc}^2 \operatorname{Im}\beta \qquad (S18)$$

Because the device is an active component, the damping can be negative and therefore $Q_{eff0}$ can be negative. Since no mechanical damping was introduced, $Q_{eff0}$ is purely a measure of the efficiency of the feedback. Equation (S18) shows that Im $\beta$ can be determined directly from the complex eigenfrequency of the FEM simulation and the value of the DC current $I_{dc}$. The DC thermal equation (S15a) can be solved separately once the DC electrical equation (S13a) has been solved. The resulting DC thermal distribution $T_{dc}$ and the displacement mode-shape $\mathbf{u}_{ac,r}$ are shown in figure S2.

## Supplementary Discussion 4: Extraction of equivalent parameters.

*a. Equations to extract the parameters.*

To quantitatively compare the value of Im $\beta$ from the FEM model with the value from the equivalent circuit shown in figure S1, the circuit parameter values are extracted from the FEM model. This is done by choosing three reference positions $\mathbf{r}_x$, $\mathbf{r}_T$ and $\mathbf{r}_v$ in the FEM geometry and imposing the condition that the variables in the equivalent circuit in figure S1 should equal the corresponding variables in the FEM solution at these points: $x(t)=u_{ac,rx}(\mathbf{r}_x)e^{i\omega t}$, $T_{ac}(t)=T_{ac,r}(\mathbf{r}_T)e^{i\omega t}$ and $v_{ac}(t)=v_{ac,r}(\mathbf{r}_v)e^{i\omega t}$, where $u_{ac,rx}$ is the x-component of the displacement vector $\mathbf{u}_{ac,r}$. The reference position $\mathbf{r}_x$ is the centre-of-mass of the rectangular mass, $\mathbf{r}_T$ is the centre of the nanobeam and $\mathbf{r}_v$ is the position of the anchor boundary that is connected to the current source $I_{dc}$. Using these variables, the piezoresistive constant was determined from the FEM solution using:

$$K_{pr} = \frac{v_{ac}}{V_{dc}(\mathbf{r}_v)x} \tag{S19}$$

To determine the thermal parameters, first the total AC resistive heating power $p_{ac}$ is determined by integrating the AC heating power density $Q_{ac}$ from equation (S14b) over the whole volume $V_{olume}$ of the resonator:

$$p_{ac} = \int_{Volume} Q_{ac} dV_{olume} \tag{S20}$$

Although the integral is taken over the whole volume of the device, the AC resistive power $Q_{ac}$ is concentrated in the nanobeam as can be seen in figure S3. This AC heating power generates thermal waves which are approximate solutions of the 1D heat equation $c_p\rho_d\partial T/\partial t=k_h\partial^2 T/\partial y^2$. The 1D solution $T(y,t)=T_{ac,r}(\mathbf{r}_T)e^{-2\pi y/\lambda_h}e^{i(\omega t-2\pi y/\lambda_h)}$ consists of waves which emanate from the center of the nanobeam $\mathbf{r}_T$ and propagate along the y-axis with a thermal wavelength $\lambda_h=(4\pi k_h/[c_p\rho_d f_{res}])^{1/2}=26$ µm and exponentially decay in amplitude. The thermal wavelength and exponential decay correspond well to the thermal FEM simulations of $T_{ac,r}$ in the oscillator as shown in figures S4 and S5.

To determine the effective thermal expansion coefficient $\alpha$ a separate mechanical FEM simulation is performed, where the displacement $x_{te}$ as a result of the thermal expansion force of the AC temperature distribution $T_{ac,r}$ is determined. From this simulation the effective thermal expansion coefficient $\alpha$ is determined as:

$$\alpha = |x_{te}/T_{ac}| \tag{S21}$$

Using the values of $p_{ac}$, $x_{te}$ and $\alpha$, the real thermal constants $R_t$ and $C_t$ can now be determined from:

$$x_{te} = \alpha \frac{p_{ac}}{i\omega C_t + 1/R_t} \tag{S22}$$

The values of the effective spring constant $k$ and mass $m$ are determined from $f_{res}=|\omega_{res}|/2\pi$ and the integral over the strain energy using the method described in reference [S3].

### b. Experimental determination of $K_{pr}$

Before determining the parameter values from the FEM simulation using the equations (S19-S22), the effective piezoresistive gauge factor $K_{pr}=r_{ac}/(R_{dc}x)$ is approximated [S4] from the transconductance $g_m$ curves in figure 2a. The displacement amplitude at resonance when the device is externally excited by an electrostatic force $F_{el.st.}$ is approximately given by $|x_{peak}|=Q_{int}|F_{el.st}|/k$. The effective spring constant $k$ was estimated with a finite element method to be $k=256$ N/m. For small displacements, the AC electrostatic force in the parallel plate approximation is given by $F_{el.st.}=(\varepsilon_0 v_{act,ac}V_{act,dc}A_{act})/g_{act}^2$. The area and gap of the actuation electrode are respectively $A_{act}=60.0\times1.5$ µm$^2$ and $g_{act}=200$ nm, and $\varepsilon_0$ is the vacuum permittivity. When the output is AC-grounded, the AC piezoresistive current is given by $i_{ac,pr}=-I_{dc}r_{ac}/R_{dc}$ and the transconductance is defined as $g_m=i_{ac,pr}/v_{act,ac}$. From the fits in figure 2a, the peak transconductance $g_{m,peak}$ and the intrinsic Q-factor $Q_{int}$ were determined. It is found that $|g_{m,peak}|/Q_{int}=5.0\times10^{-9}$ S, for $I_{dc}=0.1$ mA and $V_{act,dc}=-1.0$ V. In combination with the equations and constants above, this allows the effective piezoresistive gauge factor to be estimated: $K_{pr}=(g_{m,peak}/Q_{int})\times(kg_{act}^2/(\varepsilon_0 I_{dc}V_{act,dc}A_{act}))=-6.4\times10^5$ m$^{-1}$, where the negative sign of $K_{pr}$ was determined using the phase of $g_{m,peak}$.

*c. Parameter values*

The simulated parameter values as obtained from the FEM simulations using equations (S18-S22) are summarized in table S2 next to the measured values. The simulated resonance frequency $f_{res}$ is slightly higher than the measured value. This can partly be attributed to the use of the 2D plane-strain approximation. A simulation in the plane-stress approximation resulted in $f_{res}=1.278$ MHz, closer to the measured value. The simulated DC resistance is found from $R_{dc}=V_{dc}(\mathbf{r}_v)/I_{dc}$. It is lower than the measured value because not the full electrical geometry was simulated. A consequence of this is that the value of $K_{pr}$ is different because it depends on $R_{dc}$, via $K_{pr}x=r_{ac}/R_{dc}$. It is therefore better to compare the product $K_{pr}R_{dc}$. If the simulated parameters from table S2 are substituted into equation (S1) a value of Im $\beta = \alpha R_{dc} K_{pr} \text{Im}(1/[(1/R_t+i\omega_{res}C_t)])=58$ A$^{-2}$ is found, which is identical to the complex eigenvalue obtained using equation (S18) thus supporting the validity of equation (S1).

**II. Supplementary Figures**

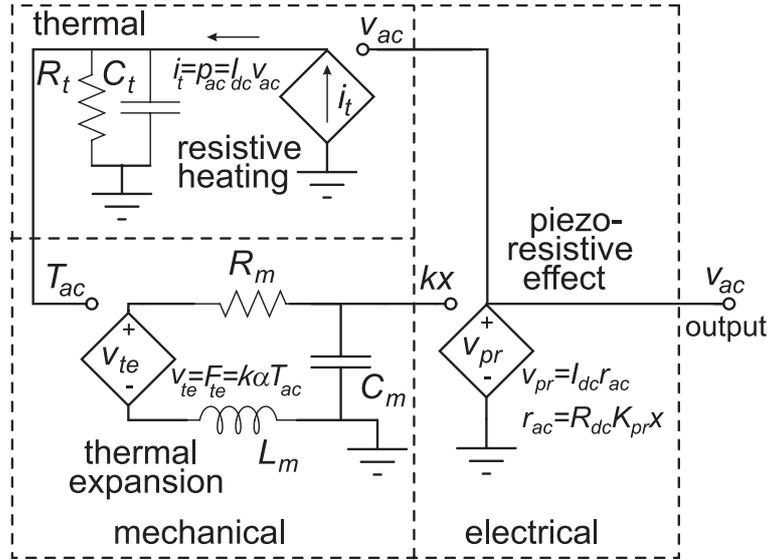

**Figure S1** Equivalent small-signal circuit model of the oscillator. In this supplementary information, this model is discussed in more detail. Moreover, finite element simulations are presented to verify the model and to extract the model parameters. This figure is identical to figure 3b in the main paper.

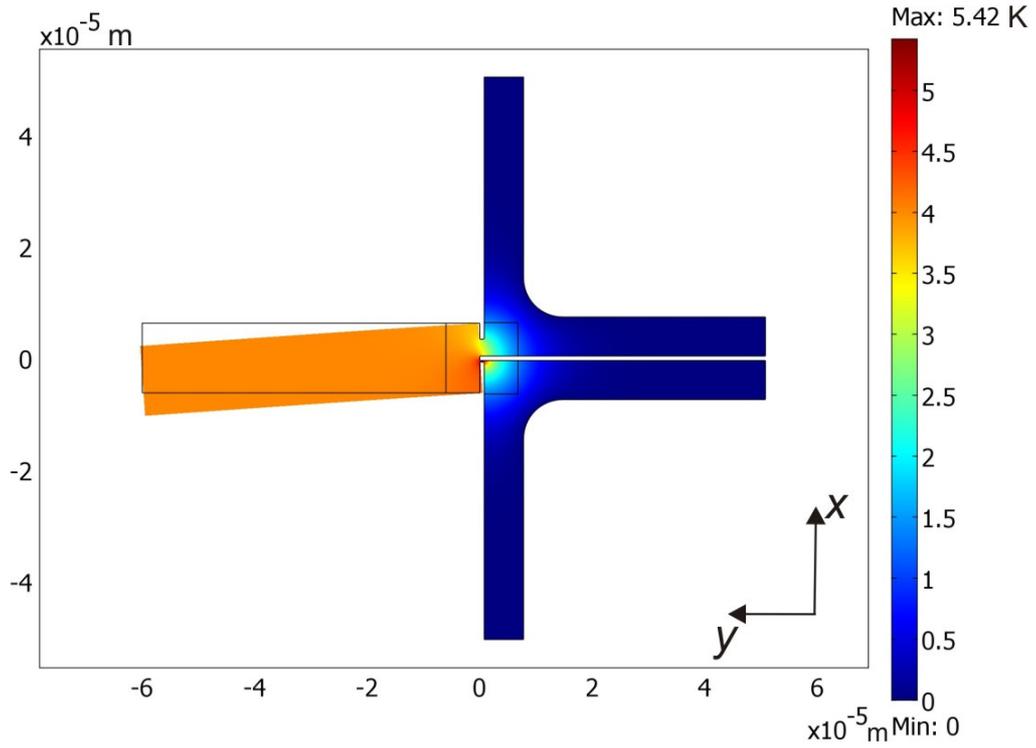

**Figure S2** Mechanical bending mode-shape $u_{ac,r}$ at maximal displacement ($x$=-2 µm) and the current induced DC temperature change $T_{dc}$ (K) of the oscillator, which is indicated by the color-scale for $I_{dc}$=1.2 mA. Because the actual amplitude of the oscillator is estimated to be only 43 nm (see legend of figure 2), the mechanical amplitude and the corresponding AC heating power and AC temperature in figures S3 and S4 are exaggerated by a factor 47.

24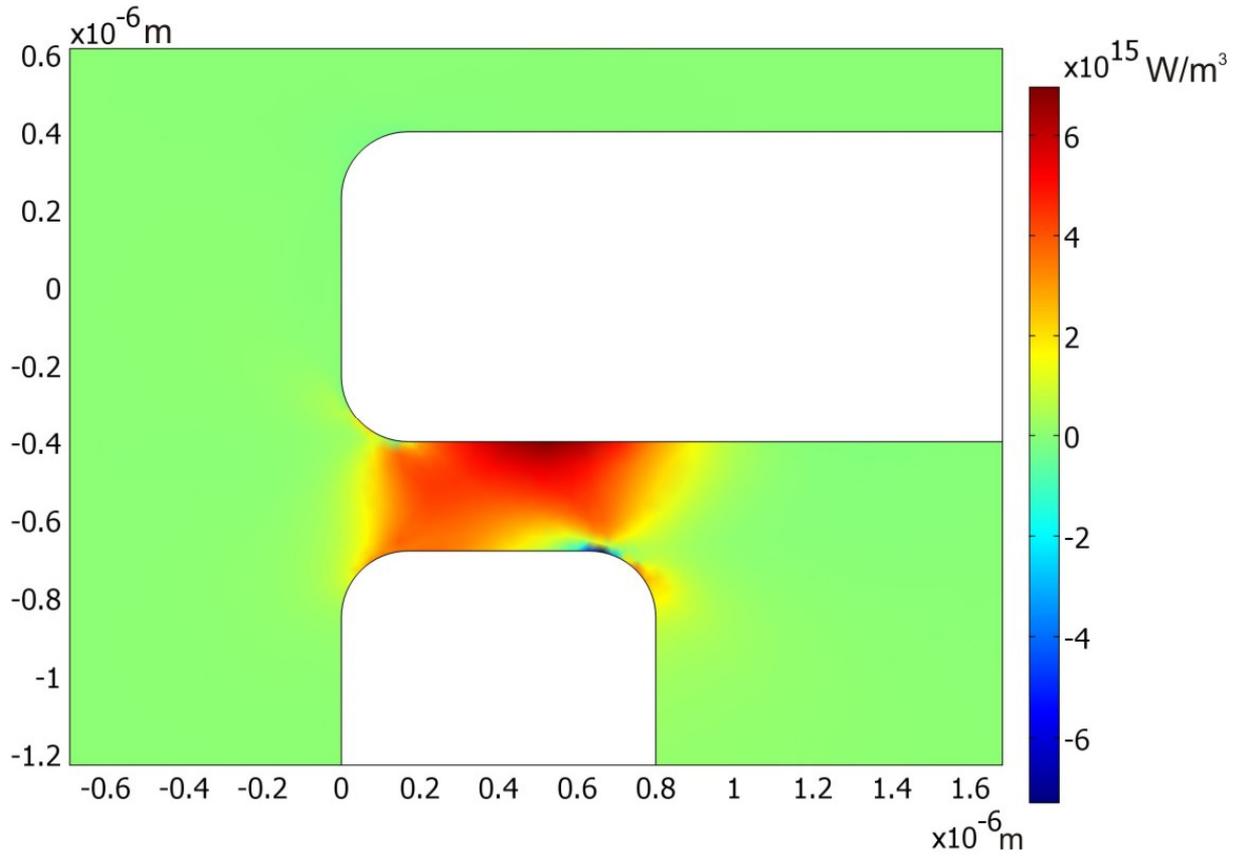

**Figure S3** AC resistive heating power density $Q_{ac}$ (W/m³) for *x*=-2 µm and $I_{dc}$=1.2 mA. Essentially all AC power is generated in the nanobeam.

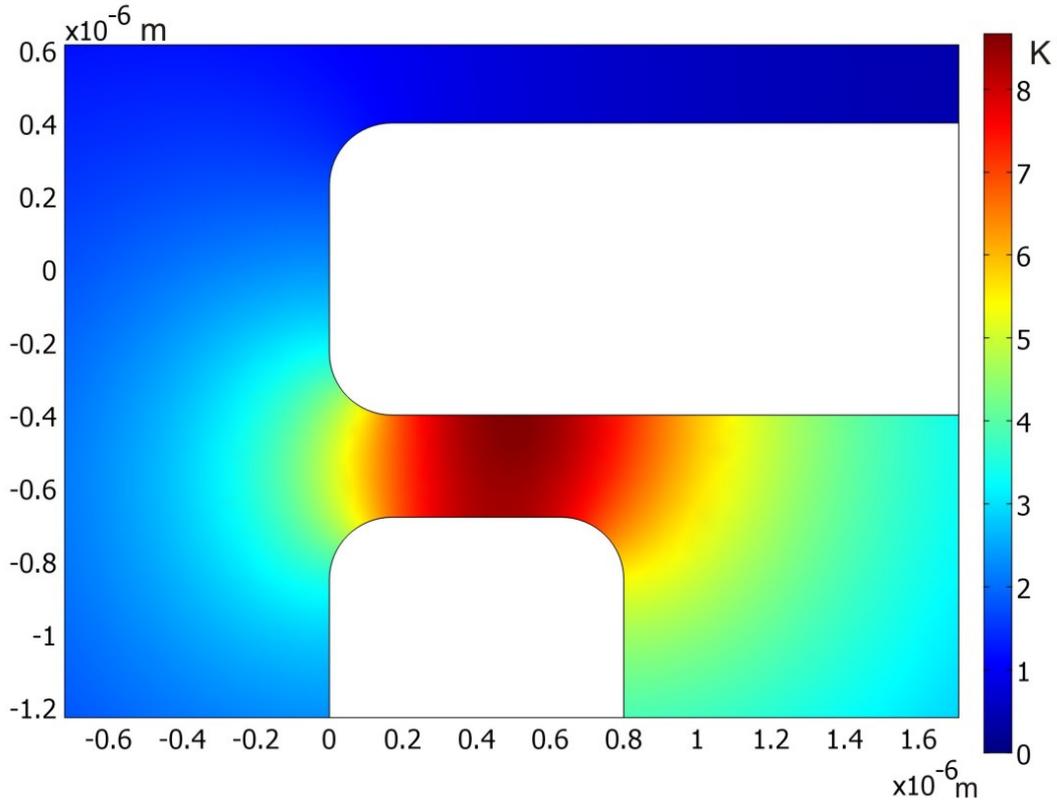

**Figure S4** The AC temperature $|T_{ac,r}|$(K) for an amplitude $x_0$=2 μm and $I_{dc}$=1.2 mA. The temperature decreases exponentially with distance from the nanobeam.

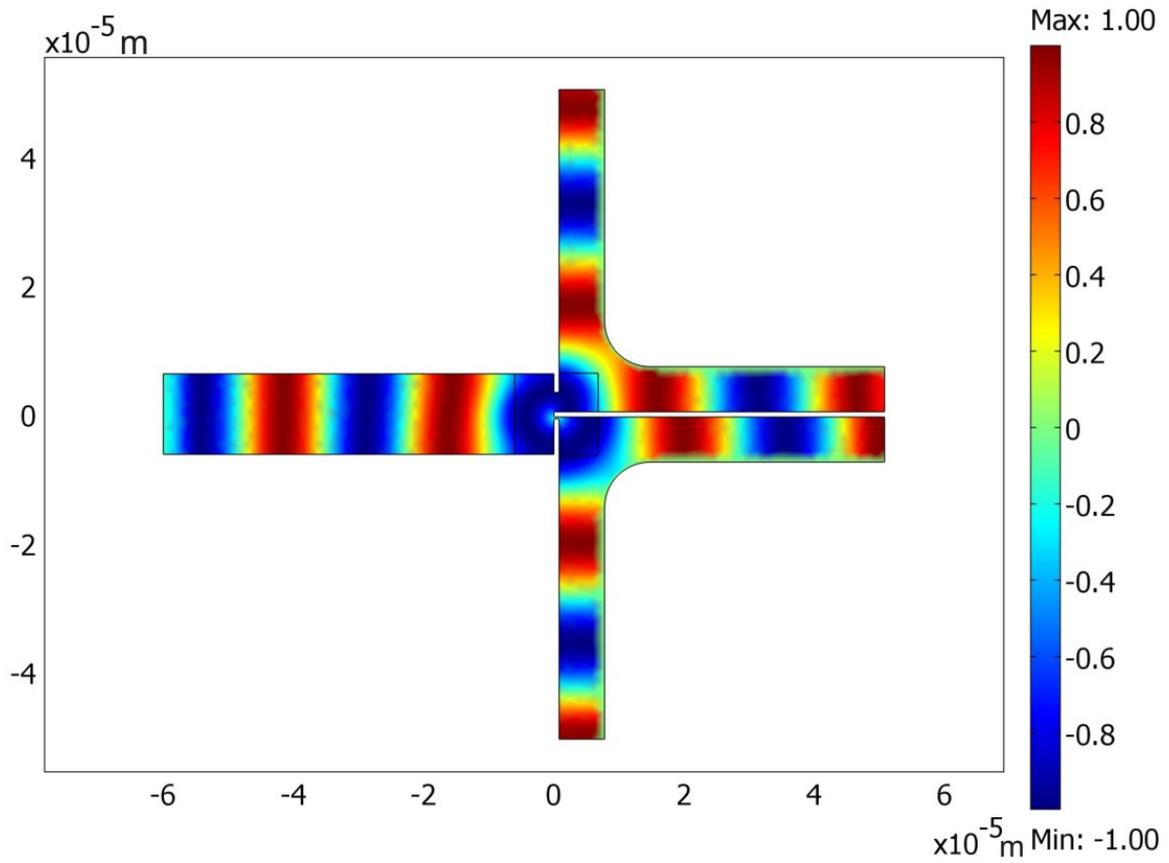

**Figure S5** Phase of the thermal waves plotted as *sin(arctan(*Im*(T$_{ac,r}$)/*Re*(T$_{ac,r}$)))*. The wavelength of the thermal waves is close to $\lambda_h$=26 μm as predicted by the 1D heat equation (see Supplementary Discussion 4a).

## III. Supplementary Tables

**Table S1 FEM material parameters at room temperature.**

| Parameter | Value | Reference |
|---|---|---|
| $c_{11}$ (GPa) | 166 | [S5] |
| $c_{12}$ (GPa) | 64 | [S5] |
| $c_{44}$ (GPa) | 80 | [S5] |
| $\rho_d$ (kg/m$^3$) | 2329 | |
| $\rho_{dc}$ (Ω·m) | $10^{-4}$ | |
| $\alpha_t$ (K$^{-1}$) | $2.6\times10^{-6}$ | [S7] |
| $\pi_{11}$ (Pa$^{-1}$) | $-102.2\times10^{-11}$ | [S1] |
| $\pi_{12}$ (Pa$^{-1}$) | $53.4\times10^{-11}$ | [S1] |
| $\pi_{44}$ (Pa$^{-1}$) | $-13.6\times10^{-11}$ | [S1] |
| $c_p$ (J/(kg·K)) | 702 | |
| $k_h$ (W/(m·K)) | 113 | [S6] |

**Table S2 Oscillator parameters.**

| | FEM Simulation | Measured |
|---|---|---|
| $f_{res}$ (MHz) | 1.335 | 1.258 |
| Im $\beta$ (A$^{-2}$) | 58 | 50 |
| $R_{dc}$ ($\Omega$) | 583 | 824 |
| $K_{pr}$ (m$^{-1}$) | $-8.35 \times 10^5$ | $-6.4 \times 10^5$ |
| $K_{pr}R_{dc}$ (M$\Omega$/m) | -487 | -527 |
| $\alpha$ (m/K) | $43.1 \times 10^{-12}$ | |
| $R_t$ (K/W) | $6.74 \times 10^3$ | |
| $C_t$ (J/K) | $9.29 \times 10^{-12}$ | |
| $k$ (N/m) | 256 | |
| $M$ (kg) | $3.64 \times 10^{-12}$ | |

## IV. Supplementary References

[S1] Smith, C. S., Piezoresistance effect in germanium and silicon. *Phys. Rev.*, **94**, 42-49, (1954).

[S2] Comsol Multiphysics, http://www.comsol.com.

[S3] Steeneken, P.G. *et al.*, Parameter extraction and support-loss in MEMS resonators. *Proc. Comsol conf.*, 725-730 (2007).

[S4] van Beek, J.T.M., Steeneken, P.G. and Giesbers, B., A 10 MHz Piezoresistive MEMS resonator with high-Q. *Proc. Int. Freq. Contr. Symp.*, 475-480 (2006).

[S5] Wortman, J.J. and Evans, R. A., Young's modulus, shear modulus, and Poisson's ratio in silicon and germanium, *J. Appl. Phys.*, **36**, 153–156, (1965).

[S6] Asheghi, M., Kurabayashi, K., Kasnavi, R., and Goodson, K. E., Thermal conduction in doped single-crystal silicon films. *J. Appl. Phys.*, **91**, 5079–5088, (2002).

[S7] Okada, Y. and Tokumaru, Y., Precise determination of lattice parameter and thermal expansion coefficient of silicon between 300 and 1500 K. *J. Appl. Phys.,* **56**, 314, (1984).